\documentclass[preprint,12pt,authoryear]{elsarticle}

\usepackage{amssymb}
\usepackage{subcaption}
\usepackage{tikz}
\usepackage{lineno}

\journal{arXiv.org}

\usepackage[pdftex,colorlinks=true]{hyperref}
\usepackage{color}
\usepackage{xcolor}
\usepackage{amsmath,amssymb}
\usepackage[]{algorithm2e}
\usepackage{graphicx}
\usepackage{thmtools}
\usepackage{float}
\usepackage{caption}
\usepackage{verbatim}
\usepackage{mathrsfs}
\usepackage{paralist}
\usepackage{mathtools}
\usepackage{wrapfig}

\usepackage{outlines}

\begin{document}


\begin{frontmatter}

\title{JetsonLEAP: a Framework to Measure Power on a
Heterogeneous System-on-a-Chip Device}

\author[label1]{Tarsila Bessa}
\author[label2]{Christopher Gull}
\author[label1]{Pedro Quint\~{a}o}
\author[label3]{Michael Frank}
\author[label2]{Jos\'{e} Nacif}
\author[label1]{Fernando Magno Quint\~{a}o Pereira}

\address[label1]{Universidade Federal de Minas Gerais, Brazil}
\address[label2]{Universidade Federal de Vi\c{c}osa, Brazil}
\address[label3]{LG Mobile Research, USA}

\begin{abstract}
Computer science marches towards energy-aware practices.
This trend impacts not only the design of computer architectures, but
also the design of programs.
However, developers still lack affordable and accurate technology to measure
energy consumption in computing systems.
The goal of this paper is to mitigate such problem.
To this end, we introduce JetsonLEAP, a framework that supports the implementation
of energy-aware programs.
JetsonLEAP consists of an embedded hardware, in our case, the Nvidia Tegra TK1
System-on-a-chip device, a circuit to control the flow of energy, of our own
design, plus a library to instrument program parts.
We discuss two different circuit setups.
The most precise setup lets us reliably measure the energy spent by 225,000
instructions,
the least precise, although more affordable setup, gives us a window of
975,000 instructions.
To probe the precision of our system, we use it in tandem with a
high-precision, high-cost acquisition system, and show that results do not
differ in any significant way from those that we get using our simpler apparatus.
Our entire infrastructure -- board, power meter and both circuits -- can
be reproduced with about \$500.00.
To demonstrate the efficacy of our framework, we have used it to measure the
energy consumed by programs running on ARM cores, on the GPU, and on a remote
server.
Furthermore, we have studied the impact of OpenACC directives
on the energy efficiency of high-performance applications.
\end{abstract}


\end{frontmatter}

\section{Introduction}
\label{sec:intro}

The efficiency of programs is usually measured in three different ways:
speed, size or energy consumption.
Presently, advances in hardware technology, coupled with new social trends, are
bestowing increasing importance upon the latter~[\cite{Sartori12}].
This importance is mostly due to two facts:
first, large scale computing -- at the data center level -- has led to the
creation of clusters that include hundreds, if not thousands, of machines.
Such clusters demand a tremendous amount of power, and ask for new ways
to manage the tradeoff between energy consumption and computing
power~[\cite{Beloglazov12}].
Second, the growing popularity of smartphones has brought in the necessity to
lengthen the battery life of portable devices.
And yet, despite this clear importance, researchers still lack precise, simple
and affordable technology to measure power consumption in computing devices.
This deficiency provides room for inaccuracies and misinformation related to
energy-aware programming techniques~[\cite{Saputra02,Valluri01,Yuki13}].

Among the sources of inaccuracies lies the ever-present question:
how to measure energy consumption in computers?
Given that the answer to such a question does not meet consensus among
researchers, conclusions drawn based on current knowledge naturally
give rise to debates.
For instance, Vetro {\em et al.}~[\cite{Vetro13}] have described a series of
patterns for the development of energy-friendly software.
However, our attempts to reproduce these patterns seem to indicate that they
are, in fact, techniques to speed-up programs; hence, the energy savings they
provide are a consequence of a faster runtime.
This strong correlation between energy consumption and execution time has
already been observed previously~[\cite{Yuki13}].
As another anecdotal case, Leal {\em et al}~[\cite{Leal16a,Leal16b}] have used
a system of image acquisition to take pictures once a second of an energy
display, in order to probe energy consumption on a smartphone.
Such creativity and perseverance would not be necessary, if they had access to
more straightforward technology.
In our opinion, such divergences happen because developers, both in the
industry and in the academia, still lack low-cost tools to measure energy
reliably in computing devices.

To remedy such omissions, this paper extends an earlier work of
ours~[\cite{Bessa16}], which introduces a precise and low-cost apparatus to
measure energy consumption in programs.
To this end, we provide an infrastructure to measure energy in a particular
embedded environment, which can be reproduced with affordable material
and straightforward programming work.
This infrastructure -- henceforth called {\em JetsonLEAP}\footnote{LEAP
(Low-Power Energy Aware Processing) is a name borrowed from
McIntire~[\cite{McIntire06}].} -- consists of an
NVIDIA Tegra TK1 board, a power meter, an electronic circuit, and
a code instrumentation library.
This library can be called directly within C/C++ programs, or indirectly via
native calls in programs written in different languages.
We claim that our framework has three virtues.
First, we measure actual -- physical -- consumption, on the device's power
supply.
Second, we can measure energy with great precision at the granularity of
about 1M instructions, e.g., 500 microseconds of execution using our
less precise circuity.
If we are allowed to use more precise equipment, we raise this accuracy to
225,000 instructions.
Contrary to other approaches, such as the {\em AtomLeap}~[\cite{Peterson11}],
this granularity does not require synchronized clocks between computing
processor and measurement device.
Finally, even though our infrastructure has been developed and demonstrated
on top of a specific device, the NVIDIA Jetson board, it can be reused with
other devices that provide general Input/Output (GPIO) ports.
This family of devices include FPGAs, audio codecs, video cards, and
embedded system such as Arduino, BeagleBone, Raspberry Pi, etc.

To demonstrate the effectiveness of our apparatus, we have used it to carry out
experiments which, by themselves, already offer interesting insights about
energy-aware programming techniques.
For instance, in Section~\ref{sec:eval} we compared the energy consumption of a
linear algebra library executing on the ARM CPUs, on the low-power ARM
core, on the Tegra GPU, or remotely, in the cloud.
We have identified clear phases in programs that perform different
tasks, such as I/O, intensive computing or multi-threaded programming.
Additionally, we have analyzed the behavior of sequential programs, written
in C, after having been ported to the GPU by means of
OpenACC~[\cite{Wienke12}] directives.
We could, during these experiments, observe situations in which the faster GPU
code was not more energy-friendly than its slower CPU version.
In short, we summarize our contributions as follows:

\begin{description}
\item [Apparatus:]
In Section~\ref{sec:sol} we explain how to build our energy measurement
infrastructure.
We believe that this description is explicit enough to enable programmers
who lack deep knowledge in electronics to reproduce our setup.
We describe two different circuits that let us switch the power
gauge on and off.
These circuits represent different tradeoffs between cost and precision.
They are built with widely available and easily affordable equipment.
Detailed manuals are also available at this project's webpage:
\url{http://cuda.dcc.ufmg.br/jetson/}.

\item [Validation:]
In Section~\ref{sub:precision} we present an empirical validation of
our methodology.
We show the precision of our equipment, and discuss threats to its
validity.
In particular, we show that more accurate machinery does not increase the
precision of our results in any meaningful way.
The recipe to reproduce these experiments is, in our opinion, one of the
core contributions of this work.

\item [Insights:]
In Section~\ref{sub:applications}, we illustrate the use of our apparatus with a
series of experiments that reveal interesting behavior of programs running on
a heterogeneous System-on-a-Chip device.
We demonstrate that it is possible to observe actual phases in the execution of
programs;
we show situations in which the fastest code is not the most energy efficient;
and we observe the power behavior of code parallelized automatically, among
other things.
\end{description}

\section{Overview}
\label{sec:ovf}

Computer programs consume energy when they execute.
Energy -- in our case electric power dissipated over a period of time -- is
measured in joules (J).
The instantaneous power consumed by any electric device is given by the
formula:
\begin{equation}
\label{eq:power}
P = V \times I
\end{equation}
Where $V$ measures the electric potential, in volts, and $I$ measures the
electric current passing through a well-known resistance.
Therefore, the energy consumed by the electric device in a given period of
time $T = e - b$ is the integral of its instantaneous consumption on $T$, e.g.:
\begin{equation}
\label{eq:energy}
E = \int_b^{e} V_f I(t)dt = V_f \int_b^{e} I(t)dt = V_f \int_b^{e} \frac{V_s(t)}{R_s}dt = \frac{V_f}{R_s} \int_b^e V_s(t)dt
\end{equation}
Above, $V_f$ is the source voltage, which is constant at the power source.
To obtain $I$ we utilize a shunt resistor of resistance $R_s$.
Thus, by measuring $V_s$ at the resistor, we get, from Ohm's Law, the value of
$I = V_s / R_s$.
One of the contributions of this work is a simple circuit of well-known $R_s$,
plus an apparatus to measure $V_s$ with high precision in very short intervals
of time.
This circuit can be combined with different hardware.
In this paper, we have coupled it with the NVIDIA TK1 Board, which we shall
describe next.

\paragraph{The NVIDIA TK1 Board}

All the measurements that we shall report in this paper have been obtained using
an NVIDIA ``Jetson TK1'' board, which contains a Tegra K1 System-on-a-chip
device, and runs Linux Ubuntu.
Tegra has been designed to support devices such as smartphones, personal digital
assistants, and mobile Internet devices.
Moreover, since its debut, this hardware has seen service in cars (Audi, Tesla
Motors), video games and high-tech domestic appliances.
We chose the Tegra as the core pillar of our energy measurement system due to
two factors: first, it has been designed with the clear goal of being
energy efficient~[\cite{Stokke15}];
second, this board gives us a heterogeneous architecture, which contains:
\begin{itemize}
\item four 32-bit ARM Cortex-A15 CPUs running at up to 2.3GHz.
\item one low-energy ARM core, which, combined with the four standard CPUs,
forms an ARM big.LITTLE design.
\item a Kepler GPU with 192 ALUs running at up to 852MHz.
\end{itemize}
Thus, this board lets us experiment with several different
techniques to carry out energy efficient compiler optimizations.
For instance, it lets us offload code to the local GPU or to a
remote server; it lets us scale frequency up and down, according
to the different phases of the program execution; it lets us
switch execution between the standard CPUs and low energy core;
and it provides us with the necessary equipment to send signals to
the energy measurement apparatus, as we shall explain in
Section~\ref{sec:sol}.

\paragraph{JetsonLEAP in one Example}

\begin{figure}[t]
\begin{center}
\includegraphics[width=1\columnwidth]{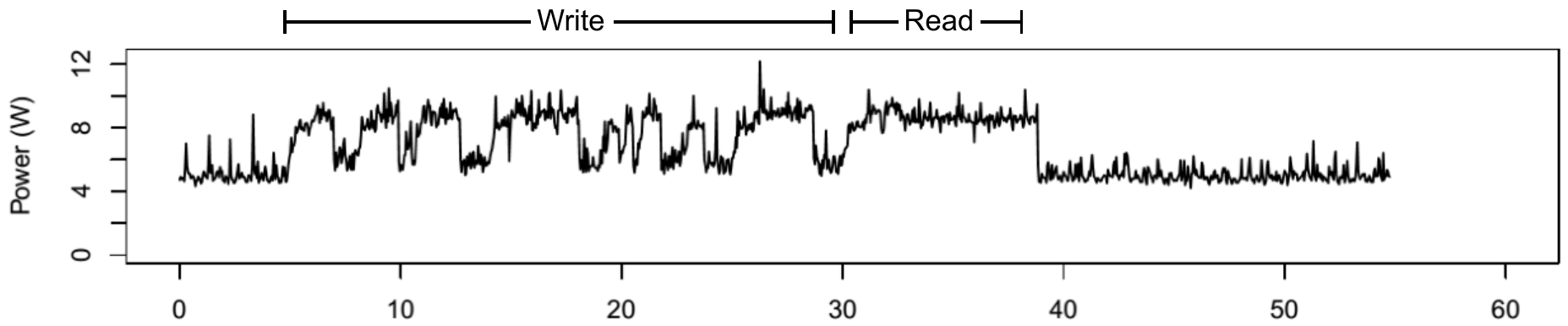}
\caption{Energy outline of a program that writes a sequence of records into a
file, and then reads them all.}
\label{fig:ReadWrite}
\end{center}
\end{figure}

The amount of energy consumed by a program is not always constant
throughout the execution of said program.
Figure~\ref{fig:ReadWrite} supports this statement with empirical
evidence. The figure shows the energy skyline of a program that
writes a large number of records into a file, and then reads this
data. The different power patterns of these two phases is clearly
visible in the figure. Bufferization causes energy to oscillate
heavily while data is being written in the file. Such oscillations
are no longer observed once we start reading the contents of that
very file. Therefore, a program may spend more or less energy,
according to the events that it produces on the hardware. This is
one of the reasons that contributes to make energy modelling a
very challenging endeavour. Hence, to perform fine-grained
analysis in programs, developers must be able to gauge the power
behavior of small events that happen during the execution of those
programs. JetsonLeap equips developers with this ability.


\begin{figure}[t]
\begin{center}
\includegraphics[width=1\columnwidth]{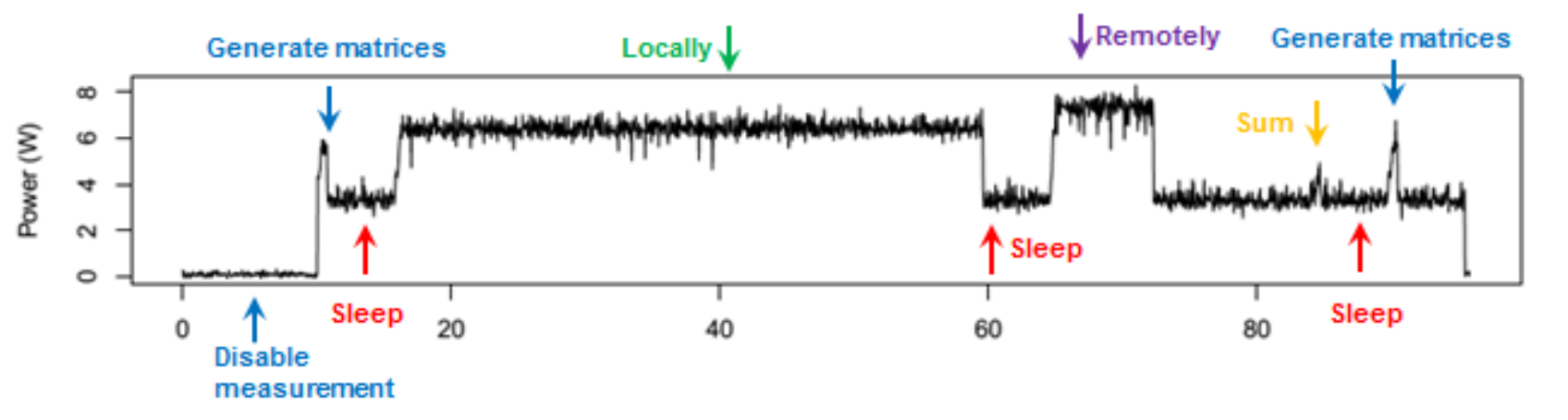}
\includegraphics[width=1\columnwidth]{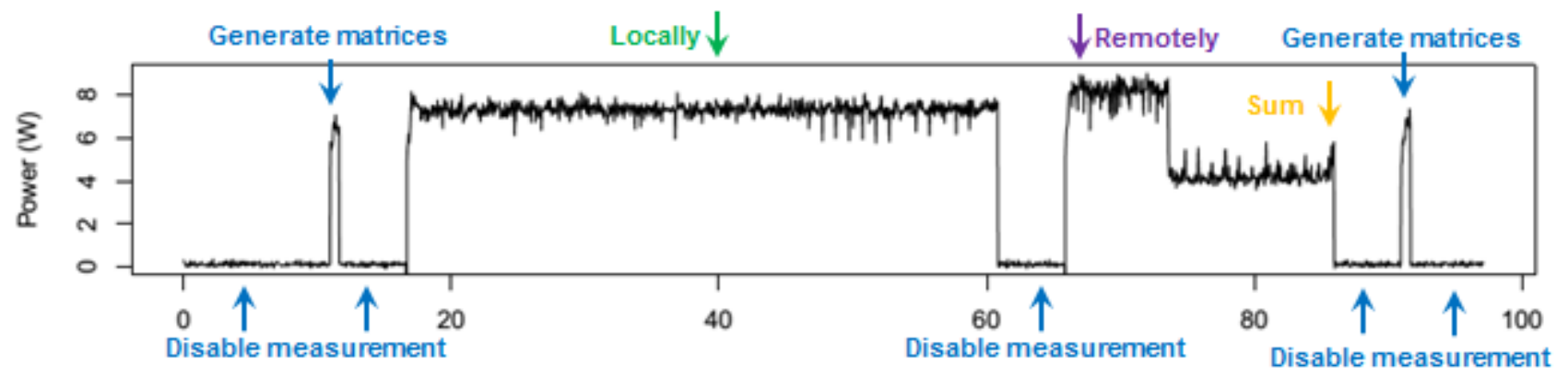}
\caption{Example showing the energy consumed at different phases of a matrix
multiplication program.
(Top) Original chart, produced without our apparatus.
(Bottom) Power chart produced with JetsonLeap.}
\label{fig:MMLocalServer}
\end{center}
\end{figure}

Figure~\ref{fig:MMLocalServer} illustrates which kind of
information we can produce with JetsonLeap. Further examples shall
be discussed in Section~\ref{sec:eval}. The figure shows a chart
that we have produced with JetsonLeap, for a program that performs
different tasks: (i) initialize two $3,000 \times 3,000$ matrices;
(ii) multiply these matrices locally; (iii) send these matrices to
a remote server; (iv) read back the product matrix, which was
constructed remotely; (v) sum up the product matrix, to check if
the result is correct; and (vi) repeat step (i). We repeat step
(i) just for consistency: to ensure that multiple occurrences of
the same event lead to very similar energy numbers. Notice that
phases (ii) and (iii) have the same goal: to obtain the matrix
that results from the multiplication of two other matrices. The
difference between them is that in the former case the
multiplication happens locally, and in the latter it happens
remotely.

We have forced the main program thread to sleep for 5 seconds in between each
task.
Using this approach, we have made the beginning and the end of each phase
of the program visually noticeable.
These marks, e.g., a 5 seconds low on the energy chart, lets us already
draw one conclusion about this setup: it is better, from an energy
perspective, to offload matrix multiplication, instead of performing it
locally on the Tegra K1.
However, this {\em modus operandi}, i.e., relying on visual identification clues to
determine program phases, is far from being ideal.
Its main shortcoming is the fact that it makes it virtually impossible to
measure the energy consumed by program events of very small duration.
We could, in principle, apply some border detection algorithm to identify
changes in the energy pattern of the program.
However, our own experience has shown that at a very low scale, border
detection becomes extremely imprecise.
One of the main contributions of this paper is to demonstrate that it is
possible to mark -- in an unambiguous way -- a specific moment in the
execution of a program to obtain its energy footprint.

The bottom part of Figure~\ref{fig:MMLocalServer} shows the same chart, this
time produced with the aid of our measurement apparatus.
We have instrumented the program to deactivate energy measurement at the
points where we call sleep.
The visual separation between the different events of interest is more
noticeable.
However, more important than relying on visual clues, our instrumentation
lets us determine precisely the points where energy numbers must be acquired.
This type of acquisition is possible even when visual clues alone are not
enough to distinguish the different events, as we shall explain in
Section~\ref{sec:sol}.

\section{Measurement Infrastructure}
\label{sec:sol}

The infrastructure of energy measurement that we provide consists of two
parts:
on the hardware side, we have an electric circuit that enables or disables the
measurement of energy, according to program signals;
on the software side, we have a library that gives developers the means to
toggle energy acquisition; plus a program that reads the output of the power
meter, and produces a report to the user.
We have experimented with two different variations of the measurement
circuit.
All these variations use the same software package to acquire energy
data.
In this section we describe each one of these elements.

\subsection{Circuit 1 -- The Relay-Based Design}
\label{sub:relay}

\begin{figure}[t]
\begin{center}
\includegraphics[width=1\columnwidth]{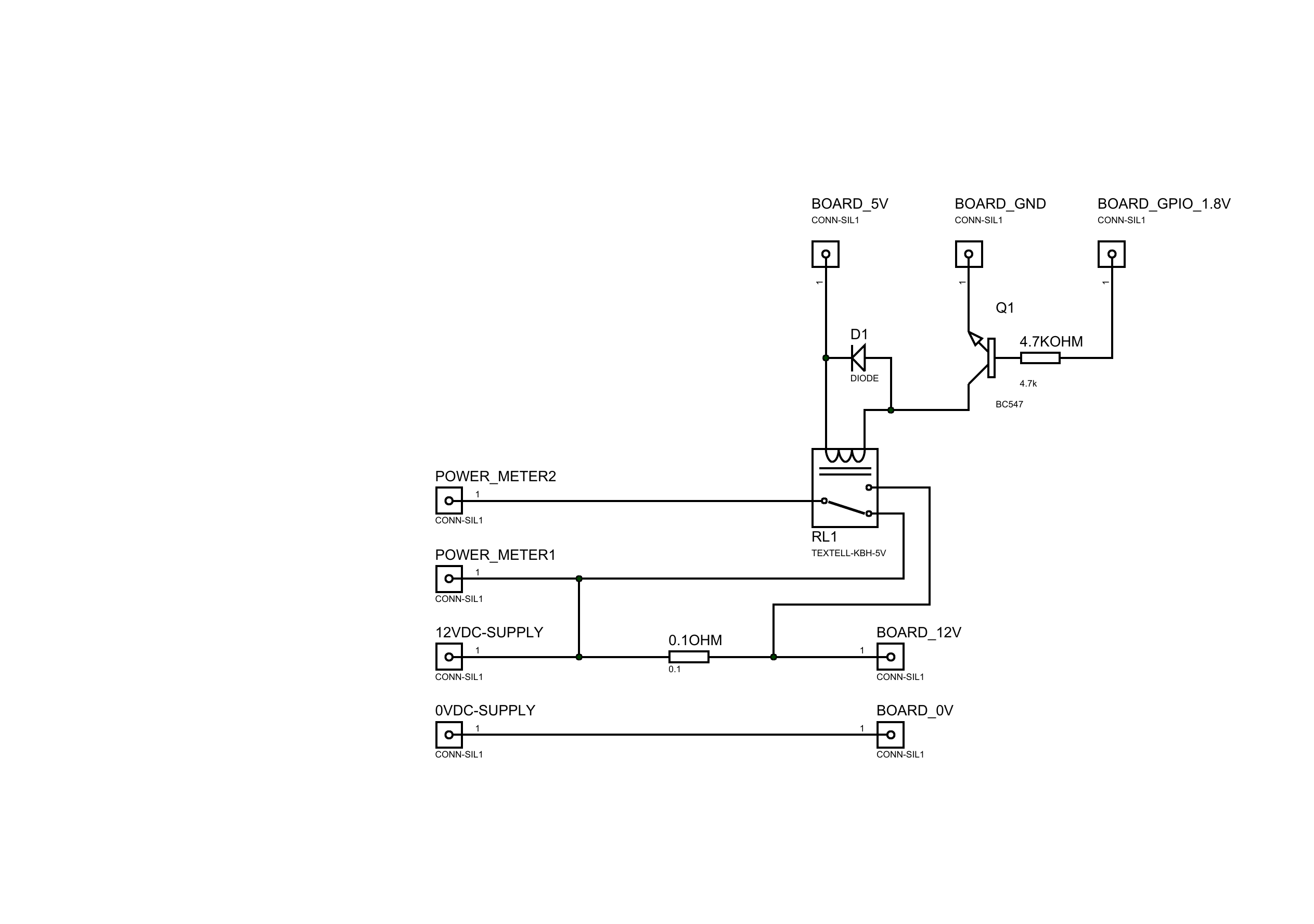}
\caption{Schematic view of the relay-based circuit that we use to measure energy
in the Jetson board.}
\label{fig:circuit}
\end{center}
\end{figure}

The first circuit that we use to gauge energy consumption uses a relay to
enable or disable measurements.
The relay is controlled by signals issued from the target program,
in such a way that only regions of interest within the code are probed.
Figure~\ref{fig:circuit} shows this design.
This apparatus consists of two sub-circuits;
one of them enables and disables the power measurements using the relay.
This relay connects the measurement probes across a shunt resistor.
The resistance is 0.1$\Omega$ at 5W, and is connected
in series with the 12V power supply input of the Jetson board. The other
circuit is responsible for actuating the relay, and consists of a 4.7k$\Omega$
resistor at 0.25W, a BC547 transistor, and a flyback diode. The trigger of the
relay is connected to a GPIO pin that in turn can be toggled by software.

\begin{figure}[t]
\begin{center}
\includegraphics[width=1\textwidth]{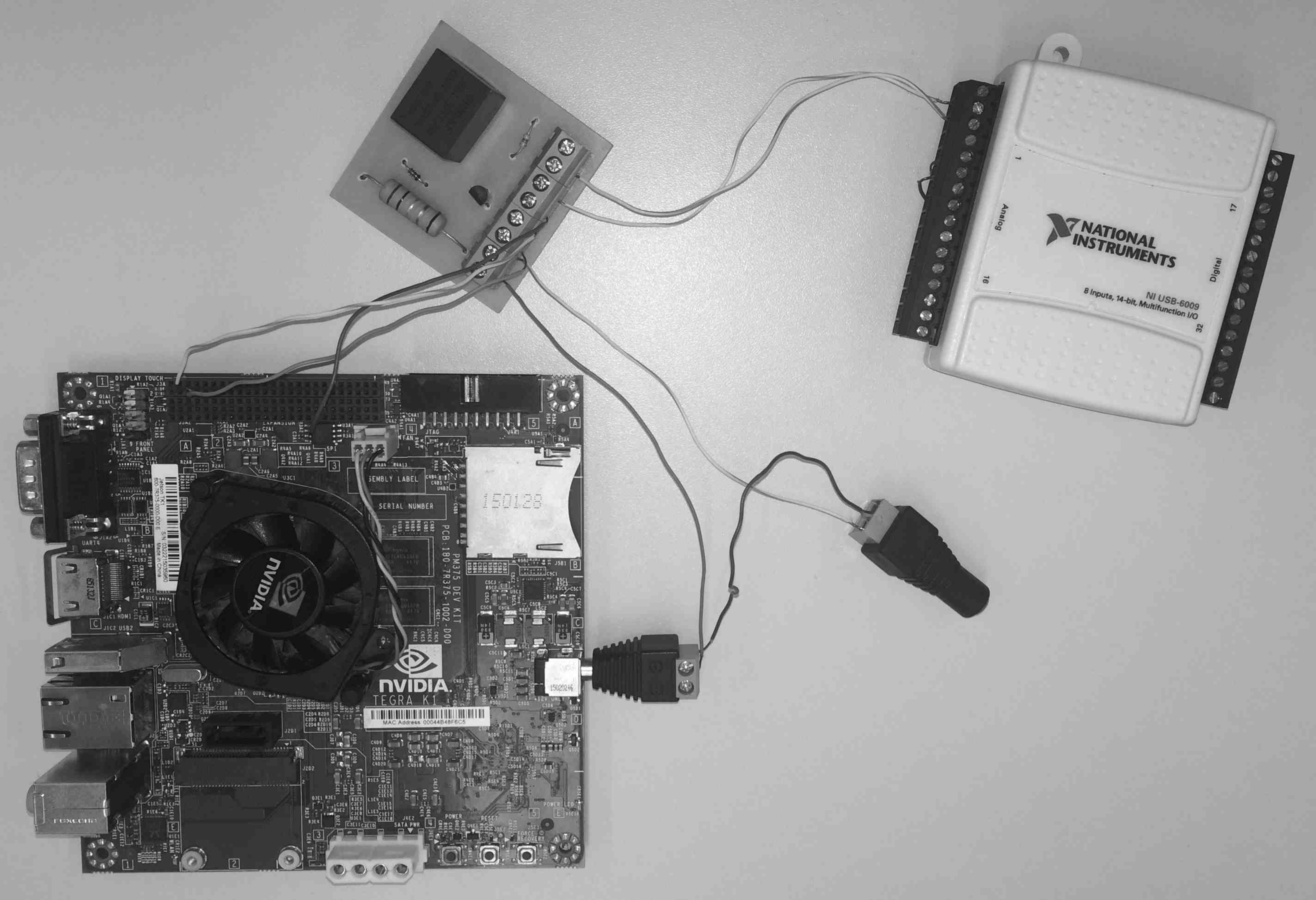}
\caption{Circuit 1 -- Relay-Based.
(Top-Left) The circuit seen in Figure~\ref{fig:circuit}.
(Bottom-Left) The Jetson board.
(Top-Right) The NI 6009 data acquisition device}
\label{fig:board}
\end{center}
\end{figure}

\begin{figure}[t!]
\begin{center}
\includegraphics[width=1\columnwidth]{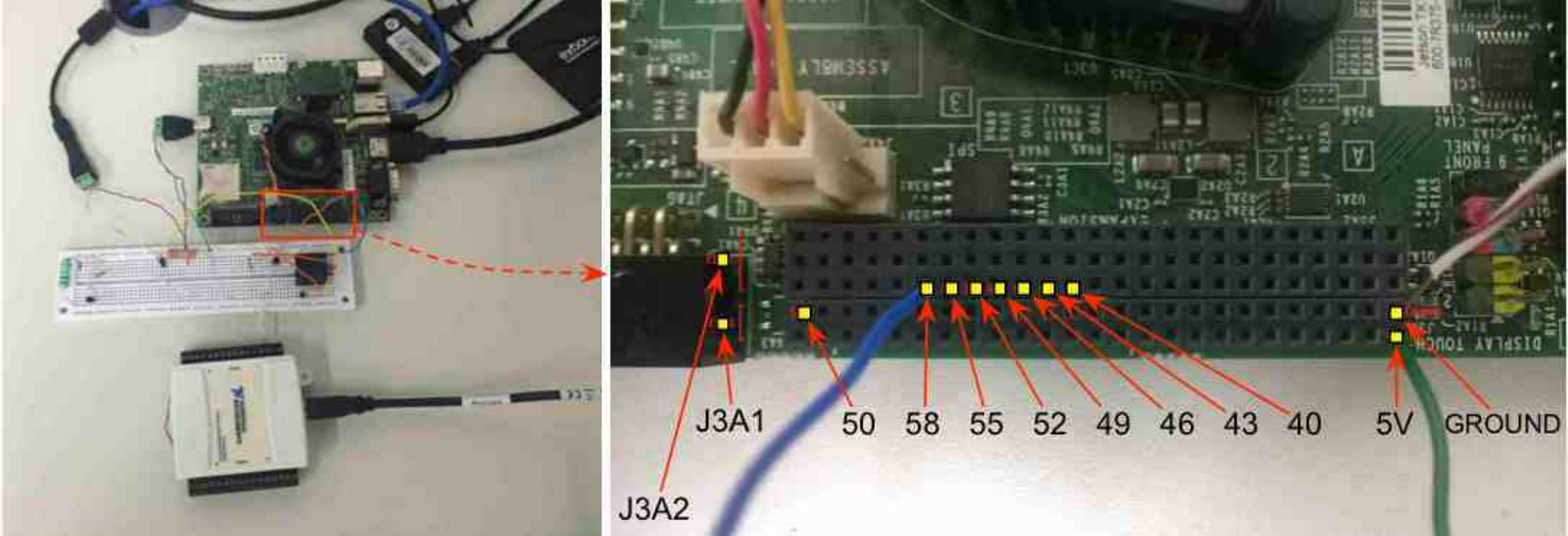}
\includegraphics[width=1\columnwidth]{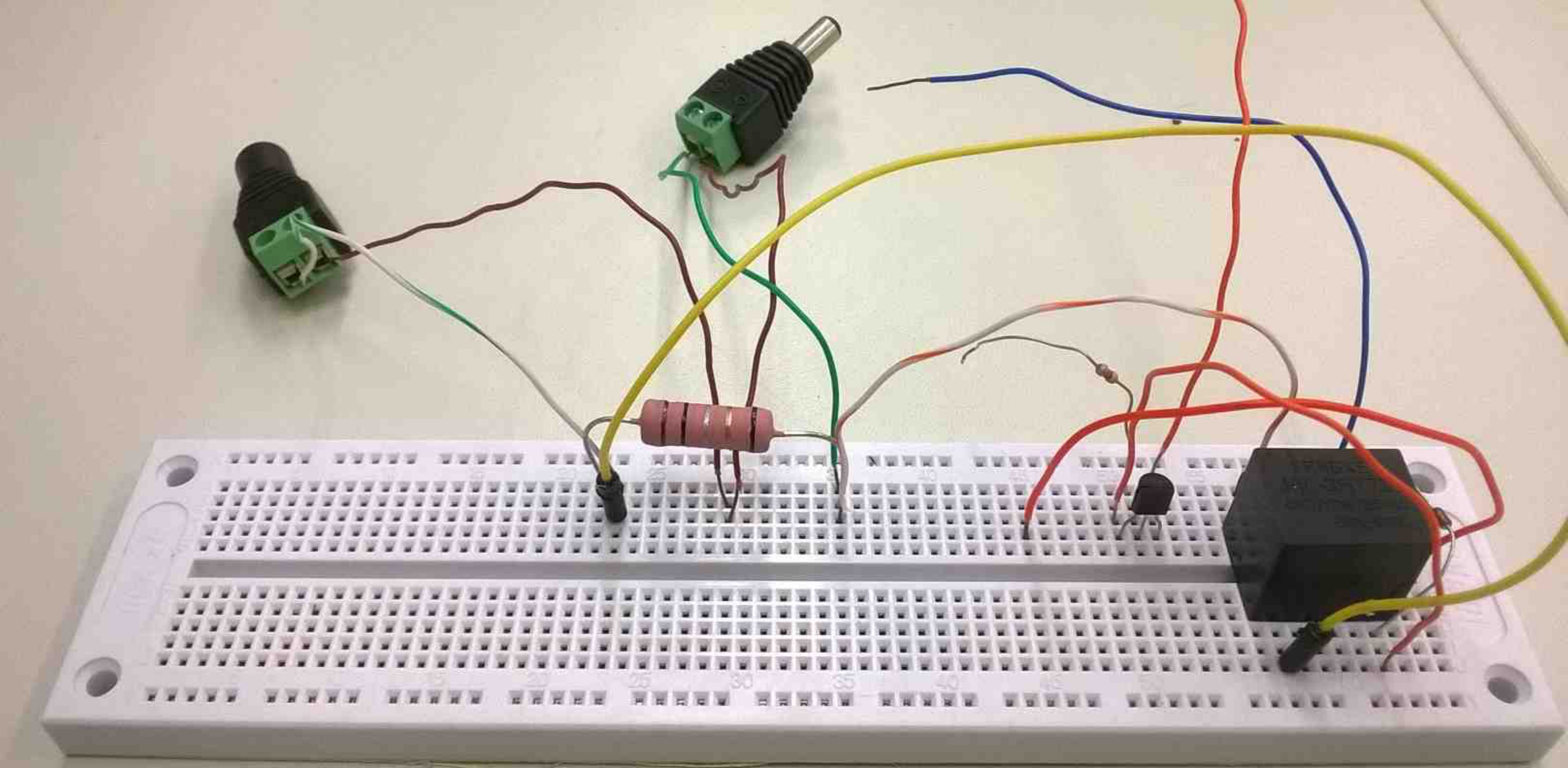}
\caption{A picture of our apparatus.
(Top-Left) The overall setup.
(Top-Right) The ports on the Jetson board.
(Bottom) Detailed view of the circuit.
We use a protoboard to make the wiring more conspicuous.}
\label{fig:Setup}
\end{center}
\end{figure}

The measurement of the power spent by the circuit is controlled by the General Purpose I/O
(GPIO) pin of the Jetson board.
The GPIO port can be activated from any software that runs on the board.
Each hardware defines GPIO ports in different ways;
in our particular case, the Jetson has eight such ports, which we have
highlighted in Figure~\ref{fig:Setup} (top-right). Additionally, the 5V supply and the ground pins can be found in aforementioned figure.
According to the Jetson's programming sheet, these ports are located on pins 40,
43, 46, 49, 52, 55, 58, 50, J3A1 and J3A2~\footnote{\url{http://elinux.org/Jetson/GPIO}}.
Each port can be signalled independently.

Figure~\ref{fig:circuit} shows that in the absence of positive signals on the
GPIO port, the two cables of the power meter perform readings at the same
logical region, which gives us a voltage of zero.
Hence, energy will be zero as well.
Upon activating a measurement --in face of a positive signal-- the transistor lets energy
flow to the relay, powering up its coil.
In this way, the relay will connect each
end of the resistor to each of the two power meter inputs,
enabling the start of the power measurement.
Using a data acquisition device (DAQ) to measure the
value of $V_s$, the difference in voltage lets us probe the
current at the shunt resistor. Using Equation~\ref{eq:energy}, this gives us
a way to know the current that flows into the Jetson board.
Figure~\ref{fig:board} shows how the circuit, the power meter and the
Jetson board are connected.

\subsection{Circuit 2 -- The Trigger-Based Design}
\label{sub:trigger}

\begin{figure}[t]
\begin{center}
\includegraphics[width=1\columnwidth]{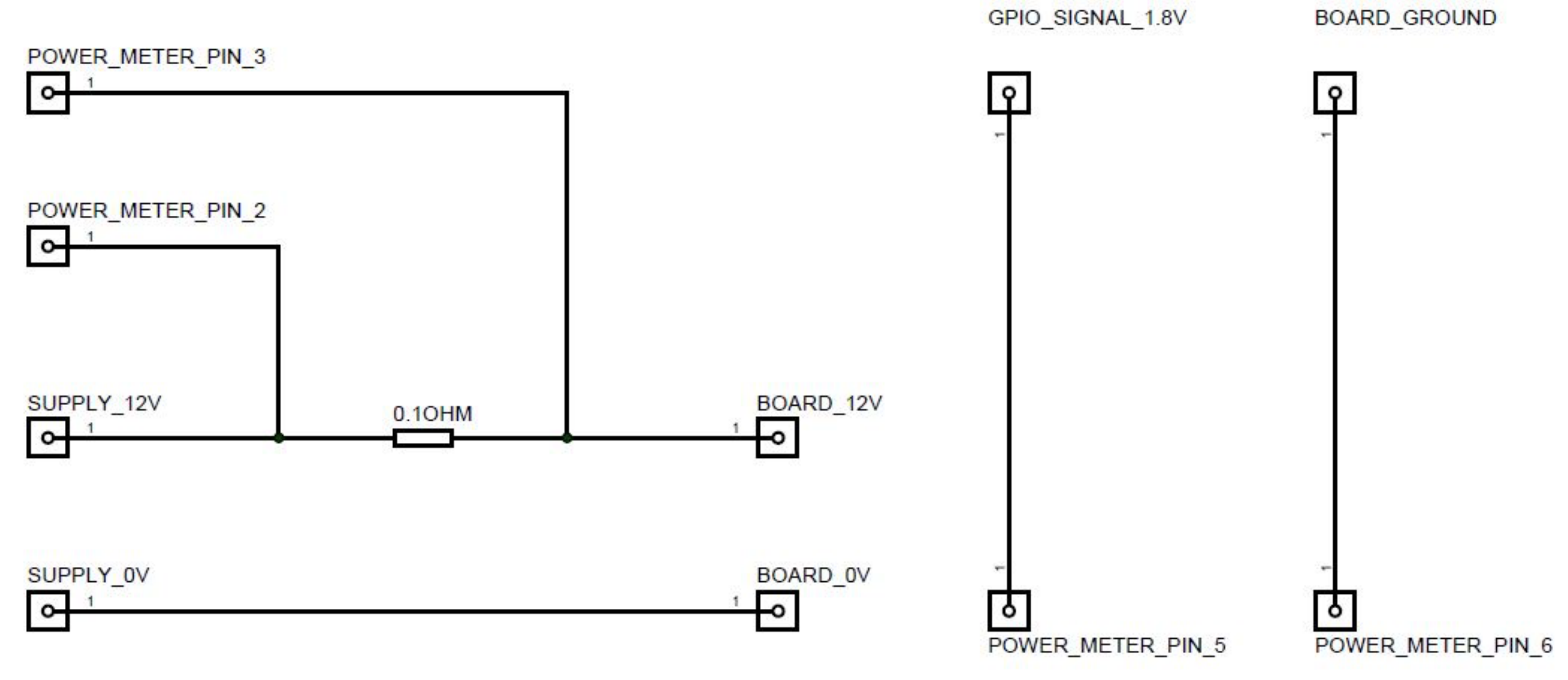}
\caption{Schematic view of the circuit that uses two channels to measure energy
in the Jetson board.}
\label{fig:trigger}
\end{center}
\end{figure}

The circuit of Section~\ref{sub:relay} lets us measure fine-grained energy
consumption events using a data acquisition device that has one probing
channel.
However, modern DAQs have more than one channel.
As an example, the NI 6009 device that we use in this paper has 16 channels.
In this case, we can replace the relay-based circuit with a much simpler design.
This new design uses an extra probing channel to read signals directly from the
GPIO port, triggering a measurement whenever the status of the port
changes.
Measurement commences once the power meter identifies a voltage drop in
the shunt resistor between the power source and the Jetson board.
Figure~\ref{fig:trigger} shows a schematic view of the trigger-based design.
This circuit contains only one electronic component: a 0.1$\Omega$ 5W resistor.

When compared to the relay-based design, this new approach has one important
advantage:
Using a second channel as a trigger has a much faster response time than using
a relay to open or close the measurement circuit.
Therefore, we can measure the energy consumption of events that take shorter
time.
On the other hand, there is a disadvantage: by using two channels, we split the
precision of the power meter in
half, because this device uses the same buffer to store data from all its
channels.
We have, however, not perceived any empirical difference between these two circuits
in practice,
because the energy behavior of the Jetson board does not show great variance
in short periods of time.


\subsection{Software}
\label{sub:soft}

The software layer of our apparatus is made of two parts.
First, we provide users with a simple library that lets them send signals to
the GPIO port of the target device.
Additionally, this library contains routines to record which ports are in use,
and to log events already performed.
Figure~\ref{fig:Precision}, in Section~\ref{sub:precision}, shows a program
that toggles the energy measurement circuit twice.
To provide users with some amount of thread-safety, we let this toggling to be
bound to identifiers.
In this way, we can ensure that only one thread has the privileges to switch
the state of the data-acquisition circuit.

The second part of our software layer is an interface with the data acquisition
tool.
We are currently using a National Instruments 6009 DAQ.
During our first toils with this device, we have been using
LabView~\footnote{\url{http://www.ni.com/labview/pt/}} to read its
output.
LabView is a development environment provided by National Instruments
itself; thus, it already comes with an interface with the DAQ.
However, for the sake of flexibility, and in hopes of porting our system to
different acquisition devices, we have coded a new interface ourselves.
Our tool, called {\em CMeasure}, has been implemented in C++.
It lets us (i) read data from the DAQ;
(ii) integrate power, to obtain energy numbers; and
(iii) produce energy reports.
Concerning (ii), while in its idle state, our circuit still lets some noise pass to the
DAQ, which oscillate between -0.001 and +0.001 watts.
The expected value of this data's integral is zero.
Thus, by simply integrating the entire range of power values that we obtain
through CMeasure, we expect to arrive at correct energy consumption with very
high confidence.

\section{Evaluation}
\label{sec:eval}

In order to validate our energy measurement system, we have used it as the
baseline platform to run different experiments.
In this section, we discuss some of the results that we have obtained in the
process.
Section~\ref{sub:precision} starts our discussion by presenting data related to
the precision of our approach.
In particular, we investigate the minimum number of instructions whose
execution we can detect using the two different circuits.
At the end of the section, we show how our setup compares against more
sophisticated equipment.
Section~\ref{sub:applications} provides material that demonstrates the many
possibilities that our platform opens up in the research community.
These experiments compare the energy footprint of
sequential, parallel (multi-core and GPU) and remote execution of programs.
We emphasize that these experiments, per se, are not a contribution of this
paper; rather, they illustrate the benefit of our framework.
Nevertheless, these experiments are original: no previous work has performed
them before on the Tegra board.

\subsection{On the Accuracy and Precision of the JetsonLeap Apparatus}
\label{sub:precision}

The goal of this section is to answer two research question:
\begin{description}
\item [Accuracy:] What is the minimum number of instructions whose energy budget
we can measure with high confidence?
\item [Precision:] How much information do we lose by using a sampling rate much
inferior to the clock of the target device we measure?
\end{description}

\begin{figure}[t]
\begin{center}
\includegraphics[width=1\columnwidth]{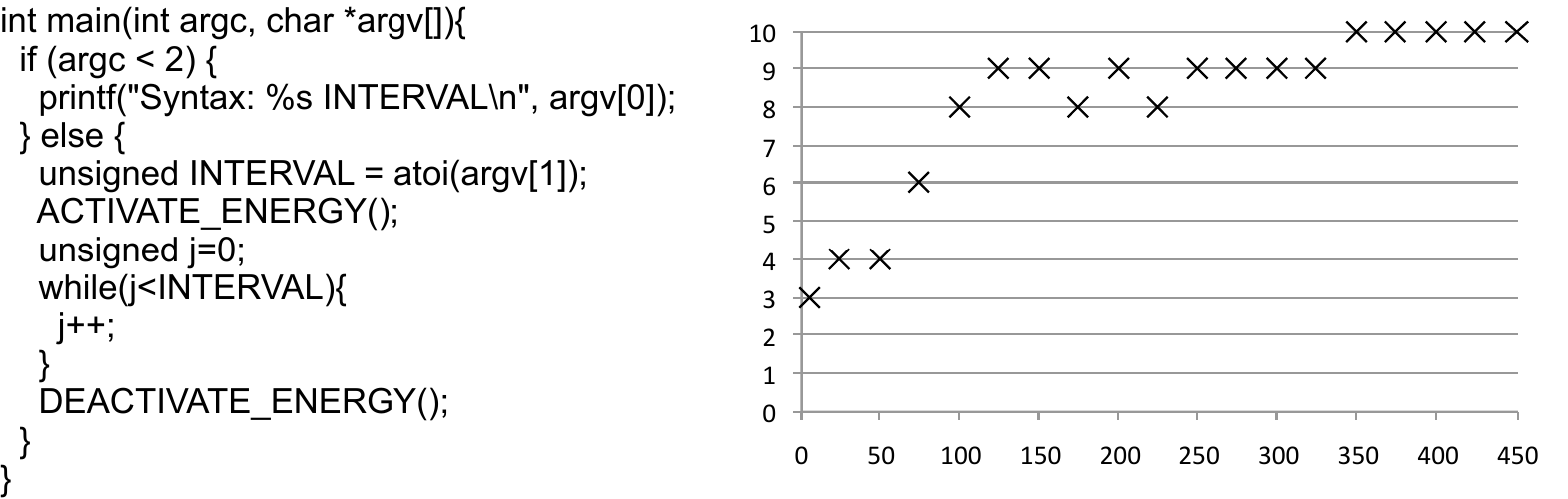}
\caption{(Left) Program used to measure the precision of our apparatus.
(Right) Chart relating the number of correct measurements with the value of
\texttt{INTERVAL} in the program on the left, using the relay-based circuit.
The Y axis gives us number of hits, out of 10 tries; the X axis gives us
the value of \texttt{INTERVAL} (in thousands).}
\label{fig:Precision}
\end{center}
\end{figure}

\paragraph{Research Question 1 -- Accuracy}
In the context of this work, an {\em event} is a sequence of
instructions processed during program execution.
An ideal measurement device should estimate with high accuracy
the power dissipated by events as small as one instruction.
However, such precision, given the high frequency of modern hardware and the
low sampling rate of data acquisition devices, is not possible.
Our first research question asks us about the minimum event size that we can
analyze with high confidence.
To produce an answer to this question, we have used the program in
Figure~\ref{fig:Precision} (left) to find out the minimum number of ARM
instructions whose energy footprint we can measure.
This program runs a loop that only increments a counter for a given
number of iterations.
By varying the number of iterations, we can estimate the minimum quantity of
instructions that gives us energy numbers that can be reproduced across
multiple experiments.
When compiled with gcc 4.2.1 -O1, the program in Figure~\ref{fig:Precision}
(left) yields a loop with three instructions: \texttt{add}, \texttt{cmp},
\texttt{blt}.
Therefore, this program gives us a rough estimate of how many instructions we
can measure: $T$ iterations yield $3T$ instructions.

Figure~\ref{fig:Precision} (right) gives us the result of this experiment
for the relay-based circuit.
For each value of \texttt{INTERVAL}, we have tried to obtain energy numbers
10 times.
Whenever we obtain a measurement, we deem it a hit; otherwise, we call it a
miss.
We know precisely if we get a hit or a miss on each sample, because we can
probe the state of the relay after we run the experiment.
We started with \texttt{INTERVAL} equals to 5,000, and then moved on to 25,000.
From there, we incremented \texttt{INTERVAL} by 25K, until reaching 450,000.
For \texttt{INTERVAL} equal to 5,000, we have been able to switch the relay
3 out of 10 times.
After we go past 325,000 iterations (975K instructions), we obtain 10 hits out of
each 10 tries.
These numbers are in accordance with the expected switching time of our
relay, i.e., half a millisecond.
Given that our ARM CPUs run at 2.3GHz during this experiment, we should expect no
more than 2.3 million instructions per millisecond.
To give the reader some perspective on the meaning of such numbers, we have
counted the total number of instructions executed with
$\mathtt{INTERVAL} = 1$.
We have removed the \texttt{ACTIVATE\_ENERGY} and the \texttt{DEACTIVATE\_ENERGY}
macros from the source code.
In this case, we have counted 7,237,290 instructions for the entire processor,
which was executed on Linux Ubuntu 14.04.
Out of this lot, 14,081 instructions execute in the \texttt{main} method alone;
and the rest are kernel instructions.
Most of these instruction are due to a loop before \texttt{main} that does a simple
search on all library functions in order to find the right ones to link dynamically.


The trigger-based circuit shows higher precision.
In this case, we obtain 10 hits out of 10 tries for
$\mathtt{INTERVAL} = 75,000$.
And for $\mathtt{INTERVAL} = 25,000$ we already obtain 8 hits out of 10 tries.
In other words, we can change the state of the power measurement circuit, with
high confidence, in intervals of $225,000$ instructions.
This accuracy is more than four times higher than that observed when using the
relay-based apparatus.
We emphasize that the sampling rate of the relay-based circuit is higher:
$40K/\mbox{sec}$, in contrast with $20K/\mbox{sec}$ when using the trigger-based
design.
As mentioned in Section~\ref{sub:trigger}, the lower sampling rate of the latter
setting is due to the fact that we must reserve half the data acquisition channel
to probe the GPIO port.
Nevertheless, even this lower sampling rate is already higher than the average
frequency used in related work.
As an example, Stokke {\em et al.} read hardware performance counters at intervals
of 100ms, in order to check the validity of their energy model for the Tegra K1
board~[\cite{Stokke15,Stokke16}].

\paragraph{Research Question 2 -- Precision}

The experiments that we perform in this paper use a relatively simple
power meter: the NI 6009 Data Acquisition Device, which has a sampling
rate of 40KHz.
If we use the trigger-based design, this sampling rate falls to 20KHz, because
we must reserve half the samples to probe the GPIO pin of the target board.
Compared to the frequency of the Jetson board, 2.3GHz, this sampling rate is
very small.
The Nyquist Theorem [\cite{Nyquist28}] states that to avoid losing any
information of an analog signal, we need to sample that signal with twice its
frequency.
If we assume that the power spent in processing instructions follows the
frequency of the processor, then our infrastructure gives us a sampling rate
much below Nyquist's rate.


In order to verify how much information we are losing, we have performed
the same measurements using a more accurate acquisition device.
We do not have access to equipment that lets us sample
energy consumption at the same rate as the frequency of the JetsonBoard,
which is 2.3GHz at maximum clock.
However, we do have access to a Keysight DSO-X 2022A Oscilloscope, whose
sampling frequency is 200MHz.
This frequency is 5,000x higher than the frequency of our NI 6009 power meter.
In this section we benchmark the power meter against this oscilloscope.
For this experiment, we used the relay-based circuit only for the power meter;
for the oscilloscope, we estimated the points at which the trigger activated
(upon program execution start) and deactivated (at execution stop).
Through this arrangement, we could calculate the energy consumed by the
Jetson board, in Joules.
The integration of the oscilloscope's instantaneous power
measurements uses the approximation rule known as the ``trapezoidal rule".

\begin{table}[t!]
    \begin{center}
        \begin{tabular}{|c|c|c|c|c|}
            \hline
            Input size & \multicolumn{2}{|c|}{Power meter} & \multicolumn{2}{|c|}{Oscilloscope}\\
            \hline
            $1,000 \times 1,000$ & 27.090 & 29.256 & 26.712 & 28.623\\
            \hline
            $1,500 \times 1,500$ & 85.647 & 85.682 & 86.597 & 95.338\\
            \hline
            $2,000 \times 2,000$ & 189.27 & 191.45 & 190.57 & 196.19\\
            \hline
            $2,500 \times 2,500$ & 375.80 & 380.42 & 373.81 & 382.81 \\
            \hline
            $3,000 \times 3,000$ & 643.15 & 654.68 & 643.32 & 652.17\\
            \hline
        \end{tabular}
    \caption{\label{tab:energycomparison} Comparison of energy consumption (in
     Joules) with Cholesky algorithm execution on the Jetson board as measured by
     the power meter and oscilloscope. We show the lower and upper boundary for the
     results obtained by each device.}
    \end{center}
\end{table}

In this experiment we use two different devices to analyze the
same Cholesky matrix factorization algorithm. The tests were
exclusively executed on one of the Jetson's CPUs, with squared
matrices of size $1,000$, $1,500$, $2,000$, $2,500$ and $3,000$
cells. Table~\ref{tab:energycomparison} shows the results of the
experimental comparisons. Results are very similar and within the margins
of error specified in Table~\ref{tab:five_oscilloscope}.
For each chosen input size, we have, thus, produced five different energy
samples in order to determine the margin of error.
Table~\ref{tab:five_oscilloscope} shows the five samples produced
using the oscilloscope.
We have interposed a waiting time between samples to
avoid overheating the processor. The perceived variation between
samples when using the power meter, was less than 1\%, given the
same input size. This variation, when using the oscilloscope, was
less than 2\%.


\begin{table}[t!]
    \begin{center}
        \begin{tabular}{|c|c|c|c|c|c|c||c|}
            \hline
            Cholesky & Test1 & Test2 & Test3 & Test4 & Test5 & Mean & ME\\
            \hline
            1000 & 26.712 & 29.644 & 27.567 & 28.623 & 27.453 & 28.000 & 1.421\\
            \hline
            1500 & 93.514 & 91.412 & 92.680 & 95.338 & 86.597 & 91.908 & 4.090\\
            \hline
            2000 & 196.19 & 192.67 & 190.57 & 193.42 & 192.79 & 193.13 & 2.507\\
            \hline
            2500 & 374.79 & 382.81 & 381.47 & 382.68 & 373.81 & 379.11 & 5.509\\
            \hline
            3000 & 643.40 & 652.17 & 645.31 & 643.32 & 649.38 & 646.71 & 4.860\\
            \hline
        \end{tabular}
    \end{center}
    \caption{Energy consumption (in Joules) with the Cholesky algorithm execution on the Jetson board as measured by the oscilloscope. Five measurements were made for each Cholesky matrix size. The Margin of Error (ME) was calculated based on all data from each of the Cholesky sizes, using a t-score based on a 95\% confidence level and degrees of freedom of 4.}.
    \label{tab:five_oscilloscope}
\end{table}

\subsection{Applications of JetsonLeap}
\label{sub:applications}

One of the contributions of this paper is to show that we can use JetsonLeap to
perform several kinds of experiments.
In the rest of this section we go over some of these experiments.

\begin{figure}[t!]
\begin{center}
\includegraphics[width=1\columnwidth]{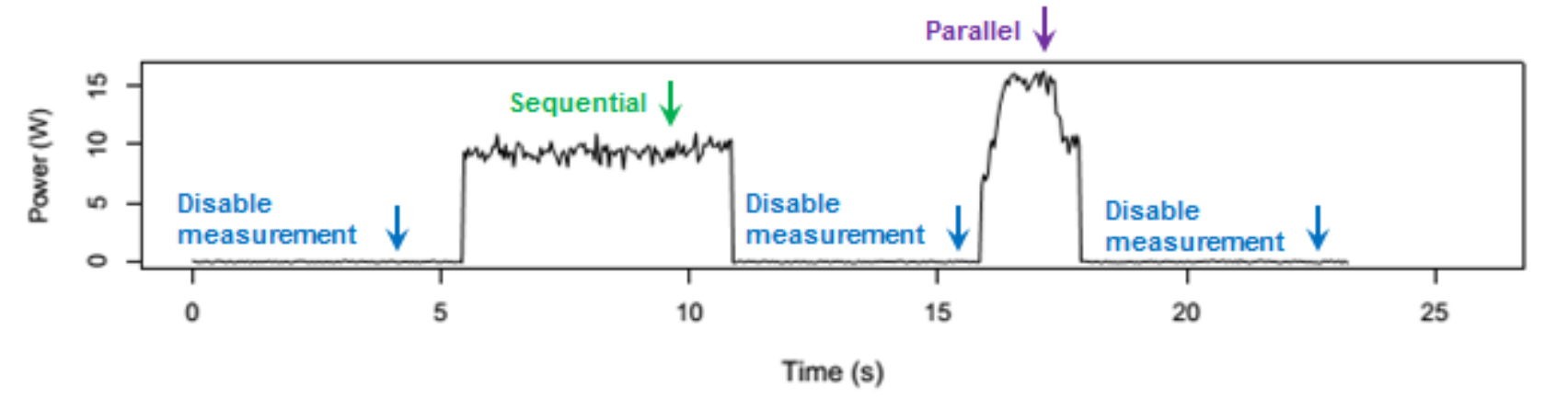}
\includegraphics[width=1\columnwidth]{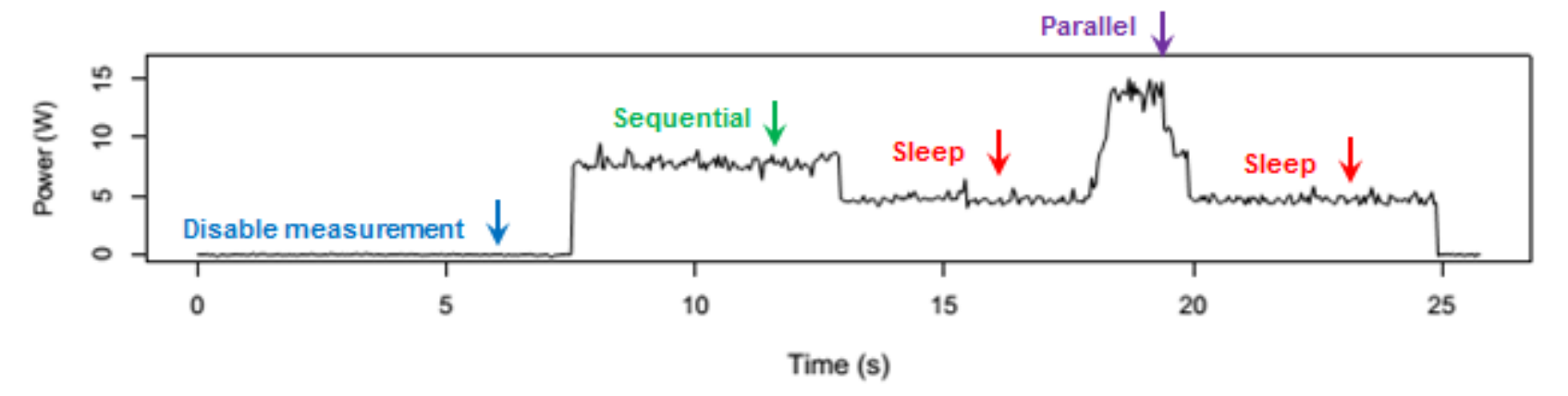}
\caption{Energy outline of a program that runs different versions of mergesort.
(Top) Chart produced with our apparatus.
(Bottom) Chart produced without energy toggling.
We use a 5-seconds sleep time to separate the execution of the different
implementations.}
\label{fig:msortsp}
\end{center}
\end{figure}

\paragraph{Parallel vs Sequential CPU code}

What is more energy efficient: to run some computation sequentially, in a single
core, or to split it into multiple cores that execute in parallel?
Different applications are likely to show different behaviors under these two
distinct circumstances.
JetsonLeap lets us probe such behavior for a particular application.
To demonstrate this possibility, we have used it to analyze the energetic
footprint of two different implementations of merge-sort: sequential and
parallel.
Both applications used in this section sort the same array of integers.
They have been compiled with gcc 4.2.1, at the -O3.
The parallel implementation of merge-sort uses Posix Threads.

\begin{figure}[t!]
\begin{center}
\includegraphics[width=1\columnwidth]{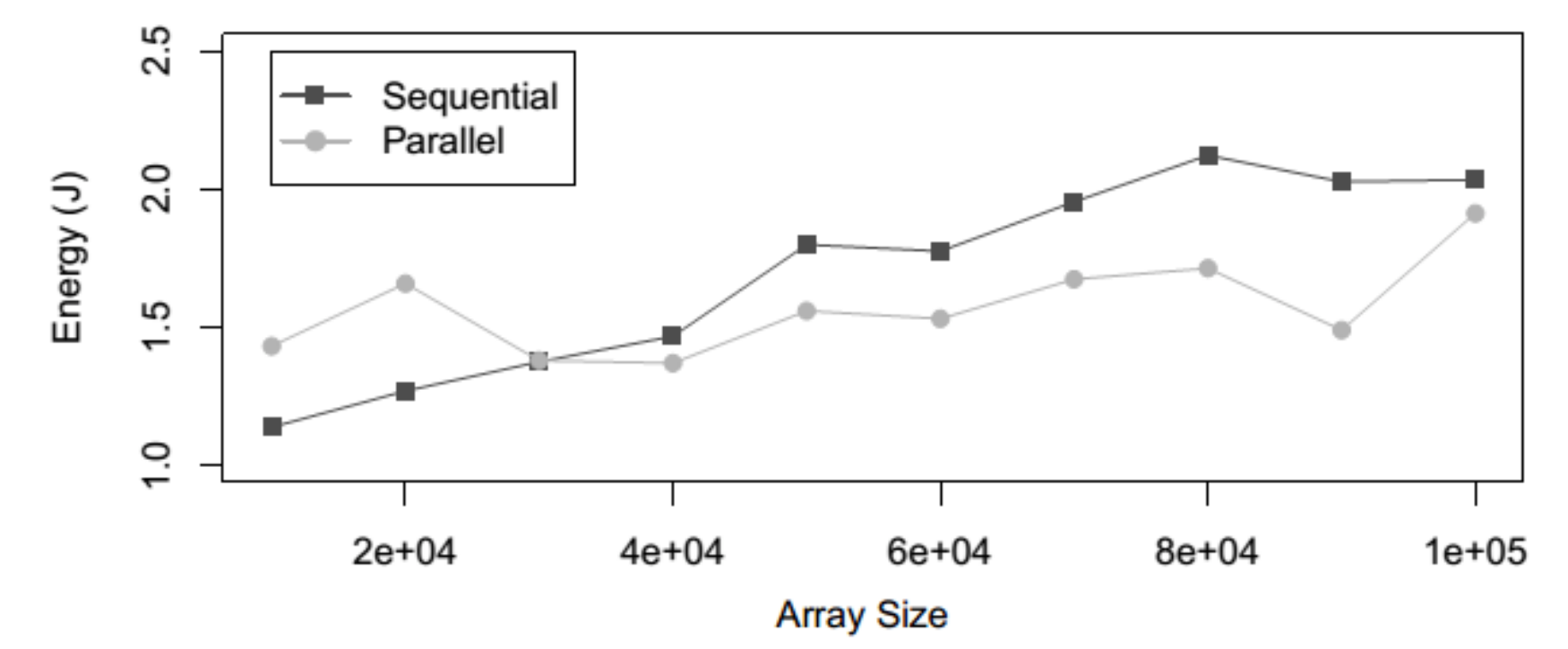}
\caption{Energy consumed by different versions of a mergesort routine.}
\label{fig:msortenergy}
\end{center}
\end{figure}

Figure~\ref{fig:msortsp} (Top) shows a power chart for the two different
implementations of merge-sort when sorting arrays of 1,000,000 integers.
The sequential implementation is slower: it takes about 5.2 seconds to sort
the input array.
Its parallel equivalent takes about 2.4 seconds to perform the same task.
However, the sequential version runs on less power: about 9W.
The parallel implementation, on the other hand, peaks at 15W.
Yet, it does not use this power at every point of its execution: as less and less
parallelism becomes available, it tends to use less threads.
This observation explains the spiky outline of the chart that JetsonLeap
produces for the parallel merge-sort.
For this specific input, an array of one million cells, the faster runtime pays
off: our parallel merge-sort uses approximately 65\% of the energy spent by the
sequential version.

\begin{figure}[t!]
\begin{center}
\includegraphics[width=1\columnwidth]{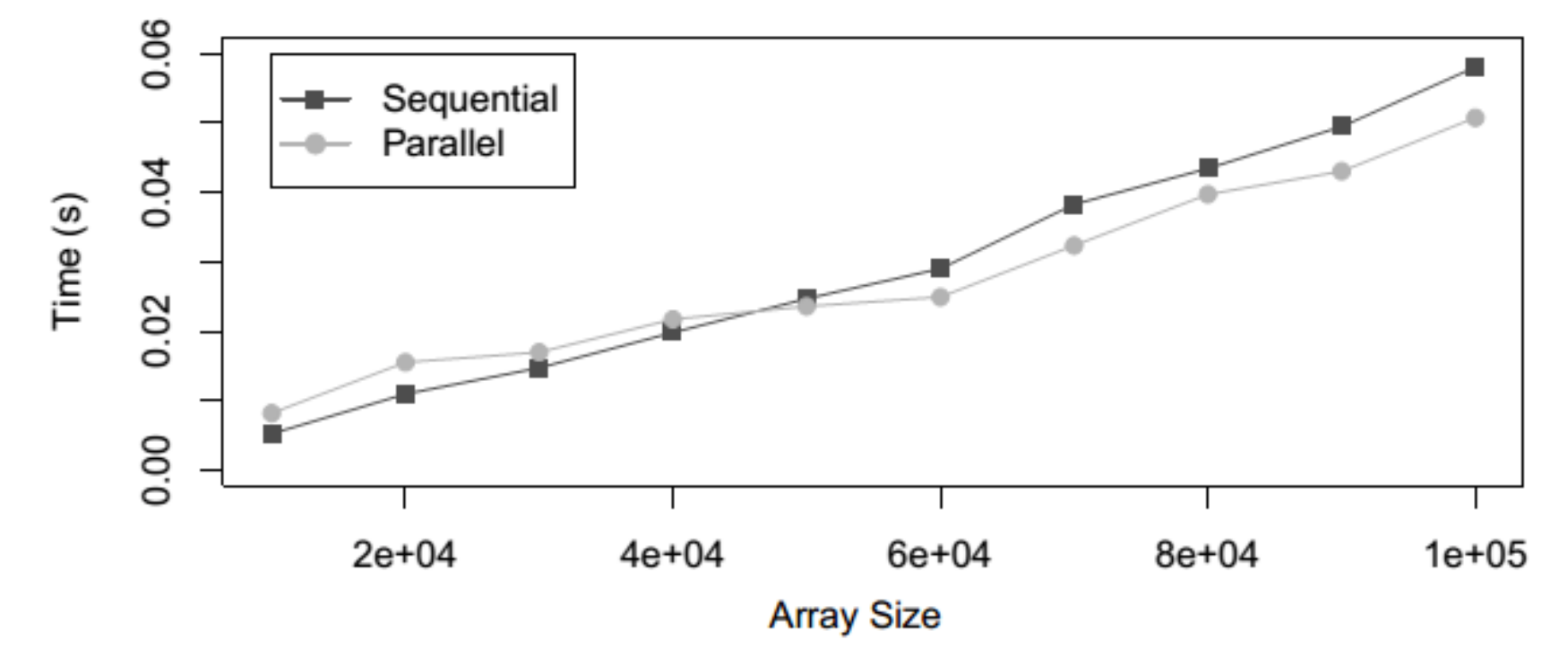}
\caption{Runtime for different versions of a mergesort routine.}
\label{fig:msortstime}
\end{center}
\end{figure}

This experiment raises another interesting question: is the parallel
implementation of merge-sort always more energy efficient than the
sequential code?
In search of answers for this question, we have fed both implementations with
arrays of different sizes.
For small inputs, the sequential algorithm runs faster, and tends to be more
energy efficient.
Figure~\ref{fig:msortenergy} shows energy consumption probed for inputs of
different sizes.
For this specific experiment, we have observed that up to arrays of 30,000 cells,
the sequential implementation uses less energy.
Past this size, the parallel version fares consistently better.
We run each sample only twice; hence, these constants may suffer small variations
in new rounds of this very experiment.

Figure~\ref{fig:msortstime} shows the runtime observed in the comparison between
the different sorting approaches with small inputs.
There is a very strong correlation between runtime and energy: the faster
implementation tends to be the more energetically efficient.
However, in this experiment we have observed a small window, extending from
arrays of 30,000 to 40,000 cells, when the faster parallel implementation
spends more energy than the slower sequential algorithm.
This behavior has been reproduced consistently in further repetitions of the
same experiment.

\paragraph{A Brief Evaluation of Code Offloading Techniques}

Code offloading consists of sending to an external host computation that the
local processor deems worthy of executing remotely.
Modern heterogeneous architectures furnish developers with a plethora of
strategies to offload code.
As an example, the TK1 board gives us the following alternatives to run
computation: the quad-core CPUs, the low-power ARM core and the GPU.
Additionally, we can offload code to a remote server, trading computation for
bandwidth.
To explore these different resources, we have used them to execute two
programs: matrix multiplication and matrix addition.
Even though conceptually very simple, these two programs present us with
diametrically opposite behaviors, as we show empirically.

\begin{figure}[t]
\begin{center}
\includegraphics[width=0.8\textwidth]{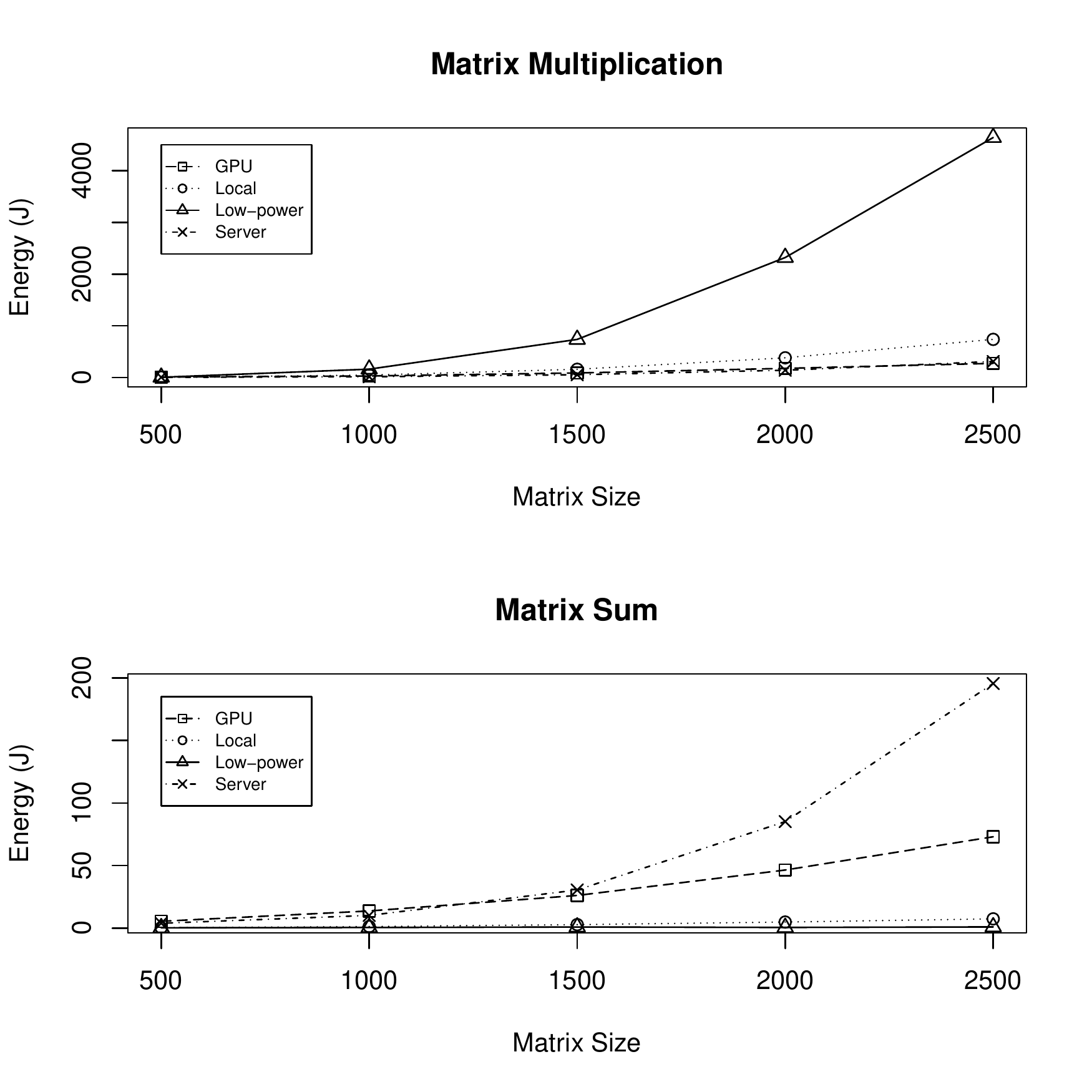}
\caption{(Top) Energetic behaviour of matrix multiplication running on
different processors.
(Bottom) Energetic behaviour of matrix addition running on
different processors.}
\label{fig:Mul-Sum}
\end{center}
\end{figure}

Figure~\ref{fig:Mul-Sum} summarizes these experiments.
The most power-efficient configuration for matrix multiplication consists of
running it on the GPU, or on the remote server.
To give the reader an idea about such tradeoffs, we spend
274.0J to multiply a $2500 \times 2500$ matrix on the GPU;
737.8J to perform the same operation on the ARM CPU, 4,639.6J if using the
low-power core, and 310.1J to offload the computation to a remote server.
Figure~\ref{fig:MMLocalServer} shows the energy outline to perform the
multiplication locally, on the standard CPU, and remotely.
In the latter scenario, all the energy spending is due to network
communication, plus idle waiting.
As Figure~\ref{fig:MMLocalServer} shows, the instantaneous power consumed in
networking is slightly higher than the power spent by CPU intensive
computations; however, the faster runtime of the server is enough to pay off
for the energy wasted with data movement.

Once we consider matrix addition we are giving a diametrically opposed
picture:
we spend 0.91J to perform the addition on the low-power core, and 7.4J to
run the same operation on the standard CPU.
Once we go to the graphics processor, this number increases tenfold: we
spent 72.9J to run matrix addition on the GPU.
Finally, if we offload the computation to a remote server, we pay a fee of
195.5J.

This remarkable difference between the two algorithms, matrix multiplication
and matrix addition, is a consequence of their asymptotic complexities:
matrix addition involves $O(N^2)$ floating-point operations on $O(N^2)$
elements of a $N \times N$ matrix.
Therefore, its computation over data ratio is $O(1)$.
Thus, the time to transfer data between devices already shadows any
gains from parallelism and offloading.
On the other hand, when it comes to the multiplication of matrices, sending the
data to a server is beneficial after a certain threshold.
Matrix multiplication has higher asymptotic complexity than matrix addition,
e.g., the former performs $O(N^3)$ floating-point operations.
Yet, the amount of data that both algorithms manipulate is still the same:
$O(N^2)$.
Thus, in the case of matrix multiplication we have a linear ratio of
computation over data, a fact that makes offloading much more advantageous.

\paragraph{Static Scheduling of Code on Heterogeneous Devices -- Manual
Annotations}

JetsonLeap allows us to compare the energy consumption of a program running
on the CPU, versus the energy consumption of similar code running on the GPU.
To demonstrate this possibility, we use a benchmark suite made of six programs,
which we took from Etino, a tool that analyzes the asymptotic complexity of
algorithms~[\cite{Demontie15}].
These programs are mostly related to linear algebra: Cholesky and LU
decomposition, matrix multiplication and matrix sum.
The other two programs are {\em Collinear List}, which finds collinear points
among a set of samples, and {\em Str. Matching}, which finds patterns
within strings.
All these are written in standard C, without any adaptations for a
Graphics Processing Unit (GPU).
To compile these programs to the Tegra's GPU, we have used
\textsf{DawnCC}~[\cite{Mendonca16}]\footnote{Available online at
\url{http://cuda.dcc.ufmg.br/dawn/}.}
to annotate them with OpenACC directives.
Each benchmark has only one core loop, which implements the bulk of the processing.
OpenACC is an annotation system that lets developers indicate to the
compiler which program parts are embarrassingly parallel, and can run on the
graphics card.
We have used accULL~[\cite{Reyes12}] to produce GPU binaries out of annotated
programs.
The code that runs on the CPU has been produced with gcc 4.2.1, at the -O3
optimization level.
Therefore, in this experiment we are comparing, in essence, the product of
different compilers, -- targeting different processors -- when given the same
source code.

\begin{figure}[t]
\begin{center}
\includegraphics[width=1\columnwidth]{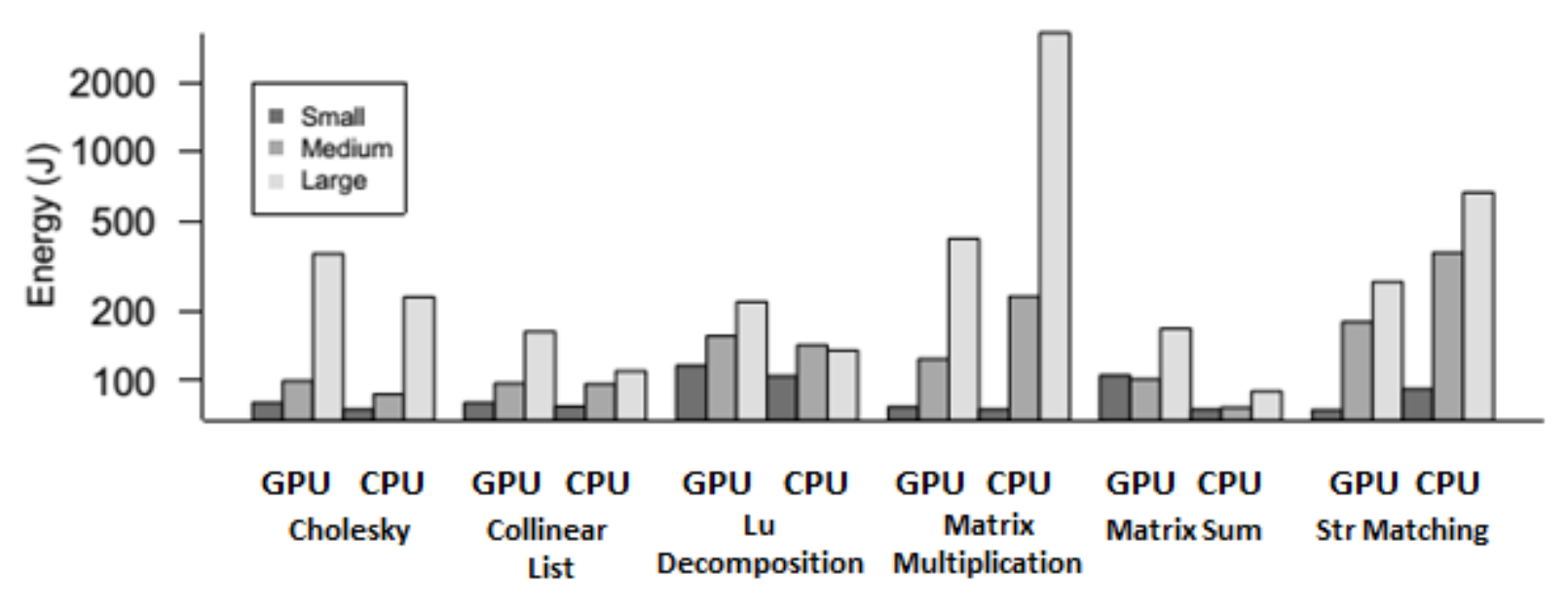}
\caption{Energy consumed by different programs, running either on the CPU, or on the GPU.}
\label{fig:energyconsump}
\end{center}
\end{figure}

\begin{figure}[t]
\begin{center}
\includegraphics[width=1\columnwidth]{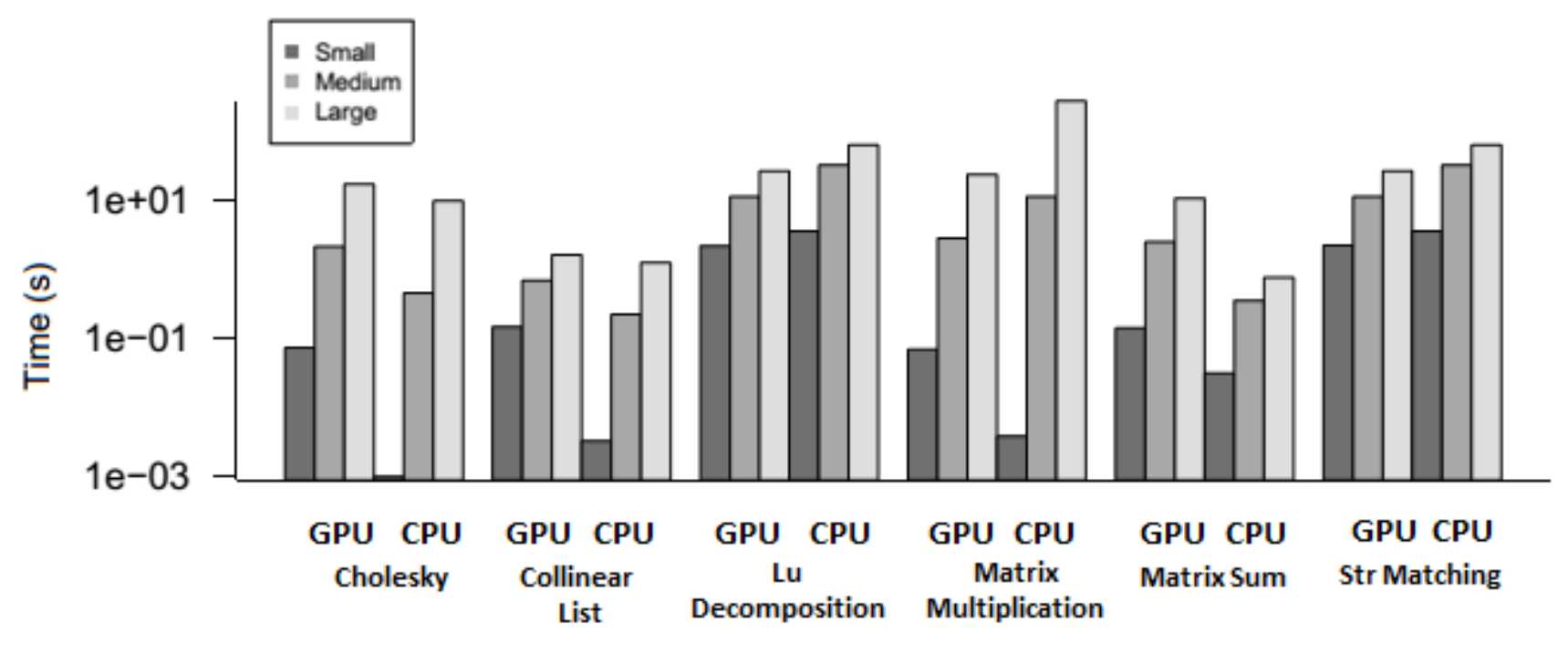}
\caption{Runtime for different programs, running either on the CPU, or on the GPU.}
\label{fig:runtime}
\end{center}
\end{figure}

Figure~\ref{fig:energyconsump} shows the energy consumed by each benchmark,
and Figure~\ref{fig:runtime} shows the time that each program takes to execute.
For each benchmark, we show results for the three different input sizes that
are available in the original distribution of Etino.
All the results that we produce for the GPU include the time (and energy) to copy
data between host (CPU) and device (GPU) processors.
We have observed that the GPU code runs faster than its CPU counterpart; however,
oftentimes this extra speed is not enough to pay for the cost of moving data.
Thus, for many benchmarks, the Tegra GPU yields worst results than the ARM
CPU.

\begin{figure}[t]
\begin{center}
\includegraphics[width=1\columnwidth]{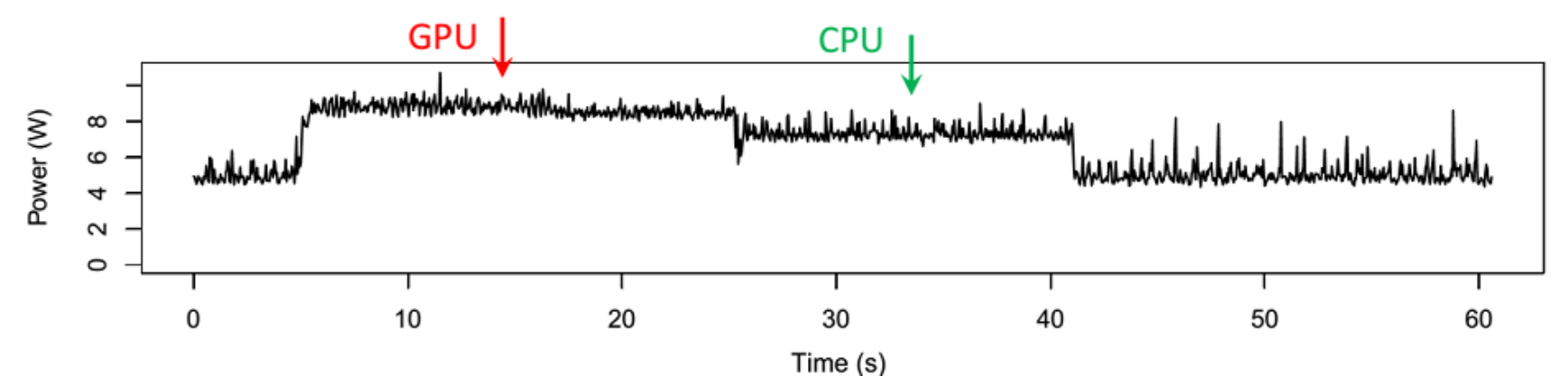}
\caption{A chart that illustrates the difference between power consumption
by a program running on the GPU and on the CPU.}
\label{fig:GPU-CPU}
\end{center}
\end{figure}

There exists a strong correlation between runtime and energy consumption; however,
there are situations in which the GPU is faster, but spends more energy.
Such fact happens twice, in \textsf{Lu Decomposition} and \textsf{String Matching}.
This result corroborates some of the conclusions drawn by Pinto
{\em et al.}~[\cite{Pinto14}], who have shown that after a certain threshold,
an excessive number of threads may be less energy efficient, even for
data-parallel applications.
Notice that they have gotten their results comparing code running on a
multi-core CPU with a different number of cores enabled each time.
Figure~\ref{fig:GPU-CPU} supports our observation.
It shows a program that performs matrix summation, first on the GPU, and then on
the CPU.
The difference in power consumption makes it easy to tell each phase apart.
During the whole execution of the GPU, its power dissipation is higher than the
CPU's.
We believe that these results are particularly interesting, because they show
very clearly that in some scenarios, runtime is not always proportional to energy
consumption.

\paragraph{Static Scheduling of Code on Heterogeneous Devices -- Automatic
Annotations}

In the previous discussion, we have manually annotated a series of simple
benchmarks with OpenACC directives, and have compared them when running on the
CPU, or on the GPU.
This time we repeat this experiment, but now using an automatic parallelizer,
\texttt{dawn-cc}~[\cite{Mendonca16}].
This tool is available through an on-line
server\footnote{\url{http://cuda.dcc.ufmg.br/dawn/}} which reads C sources, and
produces C sources annotated with either OpenACC or OpenMP directives.
The tool's distribution packs a benchmark suite, which we shall
use in this experiment.
These benchmarks are publicly
available\footnote{\url{https://cavazos-lab.github.io/PolyBench-ACC/}}.

\begin{figure}[t]
\begin{center}
\includegraphics[width=1\columnwidth]{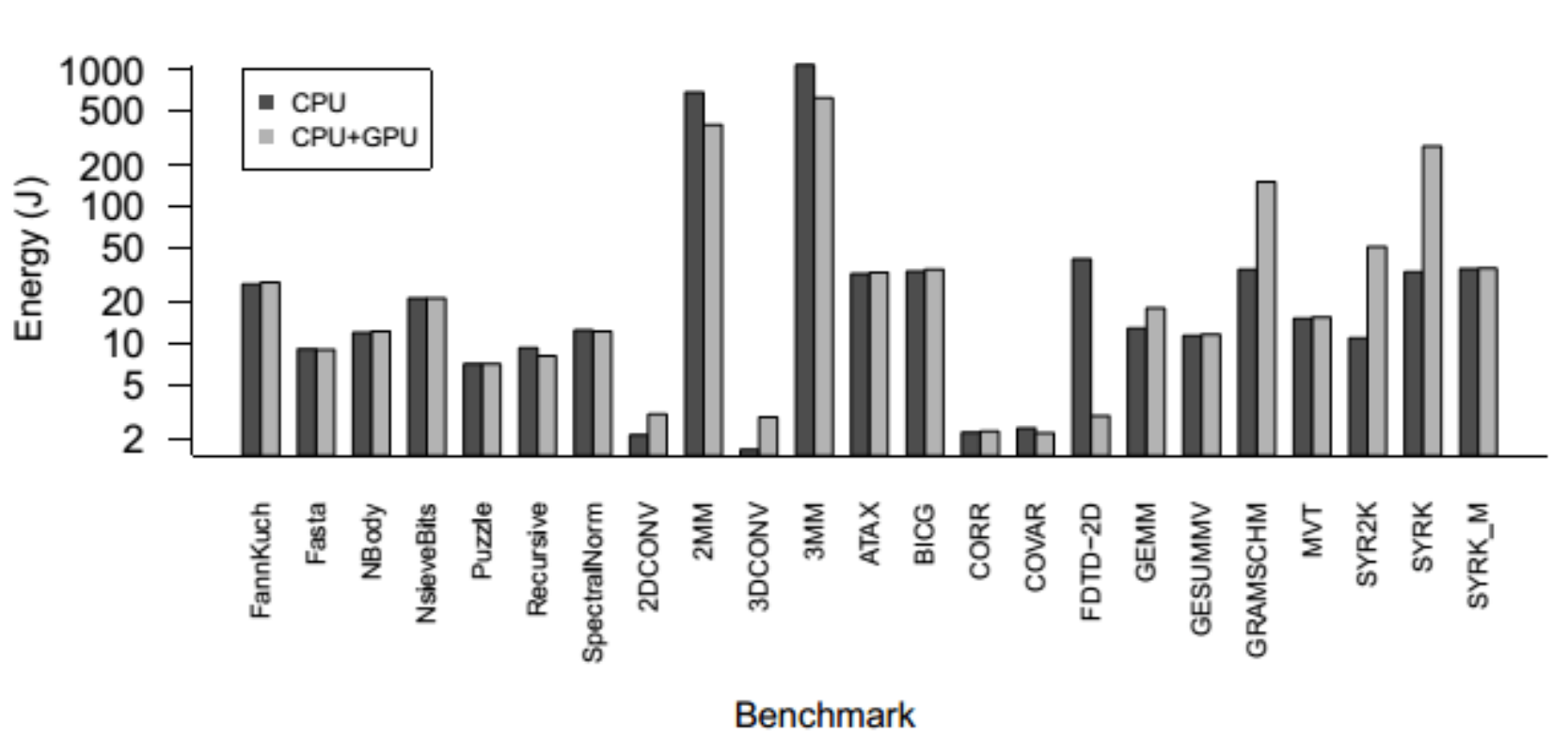}
\caption{Energy consumed by different programs, running either on the CPU, or on the CPU and GPU.}
\label{fig:cgenergy}
\end{center}
\end{figure}

For each benchmark, we run its original version on the TK1's CPU, and then feed it
to \texttt{dawn-cc}, to obtain an equivalent program annotated with OpenACC
directives.
Again, we use AccULL to compile the annotated program.
Contrary to the benchmark suite used in Figures~\ref{fig:energyconsump}
and~\ref{fig:runtime}, this new set of programs contain several loops.
The automatic annotator uses a collection of heuristics to determine the parts of
the program that must be sent to the GPU.
Annotations have been produced for all the benchmarks, except \textsf{Fannkuch},
\textsf{Fasta}, \textsf{NBody}, \textsf{NsieveBits} and \textsf{SpectralNorm}.
Nevertheless, we show the runtime and energy consumption of these benchmarks,
so that our results can be compared against the original work on
\texttt{dawn-cc}.

\begin{figure}[t]
\begin{center}
\includegraphics[width=1\columnwidth]{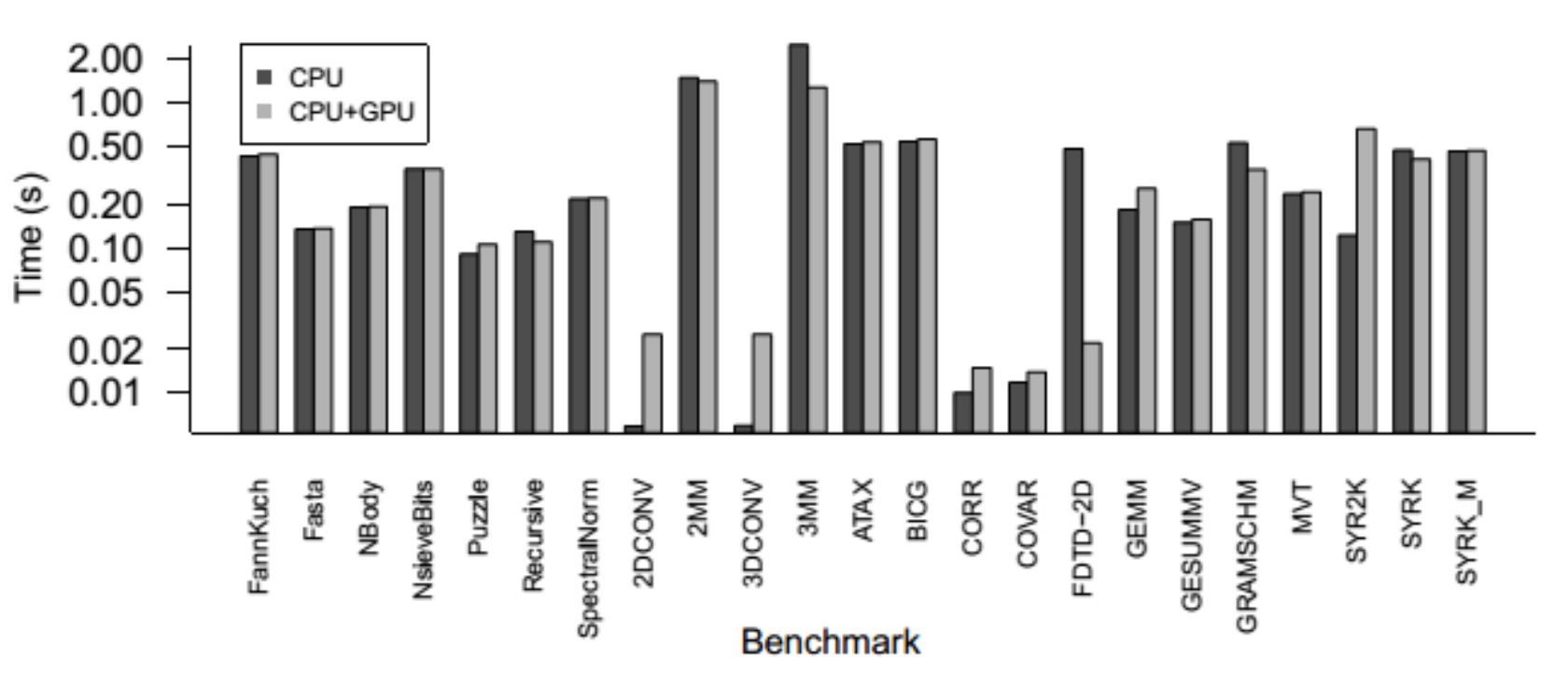}
\caption{Runtime for different programs, running either on the CPU, or on the CPU and GPU.}
\label{fig:cgtime}
\end{center}
\end{figure}

Figures~\ref{fig:cgenergy} and~\ref{fig:cgtime} summarize the results of this
experiment.
As observed with manual annotations, there is a correlation between runtime and
energy consumption.
Nevertheless, again we observe situations in which the GPU runs faster, but
spends more energy.
For instance, \textsf{GRAMSCHM}, an implementation of the
Gram–Schmidt process, we have that the GPU version is 10\% faster than the CPU's,
but consumes 30\% more energy.
We remind the reader that these numbers take into consideration the time and
energy necessary to copy data between CPU and GPU.


\section{Related Work}
\label{sec:rw}

Much has been done, recently, to enable the reliable acquisition
of power data from computing machinery. In this section we go over
a few related work, focusing on the unique characteristics of our
JetsonLeap. Before we commence our discussion, we emphasize a
point: a lot of related literature uses energy models to derive
metrics~[\cite{Dunkels07,Steinke02,Stokke16}]. Even though we do
not contest the validity of these results, we are interested in
direct energy probing. Thus, models, i.e., indirect estimation,
are not part of this survey. Nevertheless, we believe that an
infrastructure such as JetsonLeap can be used to calibrate new
analytical models.

The most direct inspiration of this work has been
AtomLeap~[\cite{Peterson11}]. Like JetsonLEAP, AtomLeap is also a
system to measure energy in a System-on-a-Chip device. However,
Singh {\em et al.} have chosen to use the Intel Atom board as
their platform of choice. Furthermore, they do not use a circuit,
like we do, to toggle energy measurement. Instead, they
synchronize the Atom's clock with a global watch used by the
energy measurement infrastructure. By logging the time when
particular events take place during the execution of a program,
they are able to estimate the amount of energy consumed during a
period of interest. They have not reported on the accuracy of this
technique, so we cannot compare it against our approach.
We believe that the Nvidia setup gives us
the opportunity to log more interesting results, given that this
hardware provides more variety of processors.
In addition to AtomLeap, our work is also related to
ARDUPOWER~[\cite{dolz2015ardupower}], which is a low-power
Wattmeter for HPC applications.
The key difference to our work is
measurement granularity, which constrains the ARDUPOWER to monitor
only power events in the range of seconds.

There is previous work that attempt to recognize programming events by means of
border detection algorithms.
This is, for instance, the approach of Silva {\em et al.}~[\cite{Silva14}], or
Nazare {\em et al}~[\cite{Nazare14}].
The idea is simple: if we assume that the hardware consumes more energy when
it runs a program, then we can expect an isolated, flat-topped hill on its
energy skyline.
Thus, the amount of energy in this clearly visible area corresponds to the
amount of energy spent by the program.
Such a methodology works to measure the energy spent by a program that runs for
a relatively long time; however, it cannot be applied to probe short
programming events, like we do in this paper.
The reasons for this limitation are two-fold.
First, internal program events might not produce visual clues that denounce
their existence.
Second, our own experience reveals that border detection requires a considerable
number of sampling points to work reliably.
This requirement would greatly reduce its precision when necessary to detect
fast events.

A final technique that is worth mentioning relies on {\em hardware
counters}, such as Intel's RAPL (Running average power limit).
Different hardware provides different kinds of performance
counters, which might log runtime, memory traffic or energy. RAPL
registers can be used to keep track of very fast programming
events, as demonstrated by H\"{a}hnel {et al}~[\cite{Hahnel12}].
Zheng {et al}~[\cite{Zheng2016accurate}] proposes a learning
algorithm that uses hardware counter information to estimate power
consumption of different platforms.
Along similar lines, Stokke {\em et al.}~[\cite{Stokke15,Stokke16}] have built
what is possibly the most precise energy model nowadays available for the TK1
board using counters.
Nevertheless, only a limited range
of computing machinery provides such tools. Thus,
techniques such as ours are still essential for
simpler hardware. Additionally, direct approaches tend to earn
more trust from the research community~[\cite{Weaver12}].

Contrary to AtomLEAP and similar approaches~[\cite{Ge10,McIntire12}], our
infrastructure does not allow us to measure the power dissipation of
separate components within the hardware, such as RAM, disks and processors.
This limitation is a consequence of the heavy integration that exists
between the many components that form the Nvidia TK1 board.
Implementing energy measurement in such environment, at component level is
outside the scope of this work.
Nevertheless, a comparison with the work of Ge {\em et al.}~[\cite{Ge10}] is
illustrative.
They use two data acquisition devices to probe different parts of the hardware
simultaneously.
Synchronisation is performed through a client-server architecture, via
time-stamps.
Although the authors have not reported the length of programming events that
they can measure, we believe that our approach enables finer measurements,
as we do not experiment network delays.
Besides, our infrastructure is cheaper: the fact that we control the
acquisition circuitry from within the target program lets us use a simpler
power meter -- even a probe with a single channel works in our case.

\section{Conclusion}
\label{sec:con}

This paper has presented JetsonLeap, an apparatus to measure energy consumption
in programs running on the Nvidia Tegra board.
JetsonLeap offers a number of advantages to developers and compiler writers,
when compared to similar alternatives.
First, it allows acquiring energy data from very brief programming events:
our experiments reveal a precision of about 225,000 instructions, given a
clock of 2.3GHz.
Such granularity enables the measurement of power-aware compiler optimizations,
for instance.
Second, our infrastructure is cheap: the entire framework can be constructed
with less than \$ 500.00, including power meter and processor.
Finally, it is general: we have built it on top of a specific platform,
the Nvidia Jetson TK1 board.
However, the only essential feature that we require on the target hardware is
the existence of a general purpose input-output port.
Such port is part of the design of several different kinds of System-on-a-Chip
devices, including open-source hardware, such as the many variations of
the Arduino single-board microcontroller.

\paragraph{Reproducibility}
Further instructions about how to reproduce our apparatus are available at
this project's webpage: \url{http://cuda.dcc.ufmg.br/jetson/}.
This site contains manuals to build the two different circuits that we use,
and to integrate them with the software stack that we have implemented.
We also provide an implementation of CMeasure, our interface with the
data acquisition device used in this work.

\paragraph{Acknowledgement}
This work have been partially funded by the Brazilian Ministry of
Science, Technology, Innovation and Communication through the National
Research Council (CNPq), by the Brazilian Ministry of Education, and by
EUBra-BIGSEA (EC Cooperation Programme
H2020 690116) and Brazilian MCTI/RNP (GA-0000000650/04).

\section*{References}

\bibliographystyle{elsarticle-harv}

\end{document}